\documentclass[showpacs,preprintnumbers,aps,prl,letterpaper,superscriptaddress,nofootinbib,tightenlines,floats,floatfix]{revtex4}
\usepackage{graphicx}
\usepackage{bm}
\usepackage{latexsym}
\usepackage{amsmath,amssymb}
\usepackage{amsfonts}
\usepackage[draft=false]{hyperref}
\usepackage[latin1]{inputenc}
\usepackage{amsthm}
\usepackage{mathrsfs}
\usepackage{ifpdf}
\usepackage{color}
\usepackage{epstopdf}
\usepackage[skip=2pt,font=scriptsize]{caption}
\usepackage{natbib}
\usepackage{blindtext}
\allowdisplaybreaks

\begin{document}
\title{Geodesic completeness of a ring wormhole}

\author{Juan Carlos Del \'Aguila}
\email{jcdelaguila@xanum.uam.mx}
\affiliation{Departamento de F\'isica, Universidad Aut\'onoma Metropolitana Iztapalapa, San Rafael Atlixco 186, CP 09340, Ciudad de M\'exico,
  M\'exico.}
\author{Tonatiuh Matos}
\email{tmatos@fis.cinvestav.mx}
 \altaffiliation{Part of the Instituto Avanzado de Cosmolog\'ia (IAC)
  collaboration http://www.iac.edu.mx/}
\affiliation{Departamento de F\'isica, Centro de Investigaci\'on y de
  Estudios Avanzados del IPN, A.P. 14-740, 07000 Ciudad de M\'exico, M\'exico.}

\begin{abstract}
In this work the possible geodesic completeness of an electromagnetic dipole wormhole is studied in detail. The space-time contains a curvature singularity and belongs to a class of solutions to the Einstein-Maxwell equations with a coupled scalar field. Specifically, a numerical analysis is performed to examine congruences of null geodesics that are directed toward the singularity. The results found here show that, depending on the strength of the coupling between the scalar and electromagnetic fields, the wormhole can be either geodesically incomplete or complete. We then focus on those wormholes that are geodesically complete and study the geometry of the neighborhood of their singularity using Kaluza-Klein theory in a five-dimensional space-time. This process allows us to provide a possible explanation of the completeness of geodesics despite unbounded curvature.  
\end{abstract}

\date{Received: date / Accepted: date}

\maketitle

\date{\today}


\maketitle

\section{I. Introduction}

Geodesic incompleteness, as a sufficient condition for a space-time to be singular, has been shown to be an useful tool for the objective of identifying any kind of singularity within a space-time \cite{wald}. A geodesic with affine parameter $\lambda$ is said to be incomplete if it is not defined for all values $\lambda\in\mathbb{R}$. In fact, the celebrated singularity theorems by Penrose and Hawking \cite{penrose, Hawking} prove the existence of incomplete geodesics if certain conditions on energy, causality, along with boundary or initial states, are satisfied. For instance the first theorem by Penrose asserts the existence of incomplete null geodesics if (a) the null energy condition is violated, (b) there is a non-compact Cauchy surface in the space-time manifold $M$ (a causality condition), and (c) there is a closed trapped surface in $M$ (a boundary condition). Penrose hence demonstrated that singularities are indeed an inherent part of General Relativity, and not simply consequences of the high degree of symmetry assumed for the solutions of the Einstein field equations. 

The answer of whether a space-time is regular or singular, though, should not be restricted to the pathological effects suffered by geodesics on the space-time manifold. Besides the paths of freely falling observers, i.e., geodesics, there may exist other physical observers within a space-time that could also be affected by a singularity. These are represented by time-like curves of bounded acceleration and are prone to incompleteness as well. If any such curve has a finite length, which is a consequence of incompleteness, then the space-time should be considered singular too. This means that, beyond being a sufficiency criterion, geodesic incompleteness is not a necessary condition to consider a space-time as singular due to the existence of various observers that can manifest irregularities. Thus, the two concepts are not equivalent. In-depth and thorough reviews on the subject can be found in \cite{Senovilla1, Senovilla2}. 

Singularities are commonly associated (but not limited) to the presence of unbounded curvature in a space-time metric. Such irregular behavior is often referred to as curvature singularity. A black hole in vacuum, as the ultimate fate of gravitational collapse, and described by the Schwarzschild metric to name an example, contains a curvature singularity in which geodesics become incomplete. In this case infinite curvature implies geodesic incompleteness. There are, however, instances of space-times that possess unbounded curvature which does not lead to geodesic incompleteness. In reference \cite{sussman}, spherically symmetric shear-free perfect fluid metrics that feature curvature divergences along complete geodesics are presented. These curves take an infinite amount of affine parameter to run into the singular regions of the space-time. Other examples of this are the wormhole geometries examined in \cite{Olmo1}. The line element is a solution to high-energy quadratic extensions of General Relativity coupled to Maxwell electrodynamics. Further study of said metric has shown that even the curves of observers with bounded acceleration are complete \cite{Olmo2}. On the other hand, the seminal two-dimensional model of the Taub-NUT vacuum space-time by Misner demonstrates that geodesic incompleteness does not imply diverging curvature \cite{taubnut}. Both notions are therefore apparently independent from each other: neither geodesic incompleteness implies the presence of curvature singularities, nor unbounded curvature implies geodesic incompleteness \cite{curiel}.

On a separate matter, but not completely unrelated, wormholes have gained the interest of physicists due to their peculiar characteristics. In particular, their ability to communicate distant regions of the same universe (or two different universes) through a kind of space-time ``shortcut'' is rather attractive. In order to facilitate traveling back and forth between both of the sides that its throat connects, it is desired for wormholes to be completely regular space-times, avoiding thus any sort of provoked ill-behavior to a physical observer. It is well-known that Morris and Thorne showed that regular, static, and spherically symmetric wormholes require exotic matter (this is, matter that violates the energy conditions) to support their throats \cite{morris}. Visser and Hochberg have extended those results, showing that the energy conditions must be violated in any regular traversable wormhole whose throat consists of a two-dimensional surface of minimal area \cite{visser1, visser2}. Nevertheless, solutions to the Einstein field equations that represent wormholes in D-dimensional gravity with a dilatonic field and singular ring bounding their throats have been alternatively found \cite{ring}. These are the so-called ring wormholes and it is possible for them to be constituted of non-exotic matter, their inconvenience is that they introduce another mayor issue: a singularity. Indeed, even before the work of Morris and Thorne, a description of the now known as Zipoy-Voorhees class of metrics, which consist of static and axially symmetric vacuum solutions of the Einstein field equations with a ring singularity, was presented in \cite{Zipoy, Voorhees}. The space-times were interpreted as having a topology that consisted of multiple connected sheets. A method to obtain those ring wormholes (and even simpler classes of them) from the Schwarzschild metric was reported in \cite{GIBBONS}. For a brief and overall review on wormholes, including the static and axially symmetric case with a ring, see \cite{bronnikovwh}. It is therefore of interest to study the implications of the singular region of this type of intriguing wormhole geometries.

In this paper we provide an example of a ring wormhole with a curvature singularity in which, depending on the value of one of its metric parameters, the space-time can be either geodesically incomplete or complete. The work is organized as follows. In section II we give a brief summary of the wormhole itself, some characteristics not reported so far in the literature will also be given. Section III consists of a description of the geodesics on the wormhole. Here, two particular simple cases of motion are studied analytically and afterward, more general geodesics are obtained numerically. We then focus on the geodesically complete type of wormhole and attempt to examine its singularity as seen in the five-dimensional space-time generated by Kaluza-Klein theory. This is done in section IV where additionally, based on the results found in the higher-dimensional analysis, an interpretation of the completeness of its geodesics is elaborated. Conclusions are drawn in the final section of the work.

\section{II. Overview of the Wormhole}
\label{s:2}
 
Our purpose in this section will be to introduce the wormhole space-time along with its general properties. The solution was previously found in \cite{DelAguila}. Here we provide a summary of its basic features and complement some physical properties that were omitted in the past reference.

The wormhole belongs to the class of Einstein-Maxwell scalar fields first described in \cite{Matos2010}. They are stationary, axially symmetric, and have a rotation parameter that makes them non-static. The general Lagrangian of the solution is

\begin{equation}
\mathscr{L}=R-2\varepsilon\nabla_\mu\Phi\nabla^{\mu}\Phi-e^{-2\alpha\Phi}F_{\mu\nu}F^{\mu\nu},
\label{eq:lagrangian}
\end{equation}
where $R$ is the Ricci scalar, $F_{\mu\nu}$ is the electromagnetic field tensor, and $\Phi$ the scalar field of a zero spin (composed) particle. Here, we set the quantity $\varepsilon=+1$ for a dilatonic field and $\varepsilon=-1$ for a phantom (or ghost) field, both with coupling constant $\alpha$. The Einstein-Maxwell-Dilaton field equations from Lagrangian (\ref{eq:lagrangian}) are

\begin{align}
R_{\mu\nu}=&2\varepsilon\nabla_\mu\Phi\nabla_\nu\Phi+2e^{-2\alpha\Phi}\left( F_{\mu\rho}F_{\nu}^{\,\,\rho}-\frac{1}{4}g_{\mu \nu}F_{\delta\gamma}F^{\delta\gamma}\right), 
\nonumber\\
\nabla_\mu(&e^{-2\alpha\Phi}F^{\mu\nu})=0, \hspace{4mm} \text{and} \hspace{4mm} \nabla^\mu\nabla_\mu\Phi=\frac{\alpha}{2}e^{-2\alpha\Phi}F_{\delta\gamma}F^{\delta\gamma}.
\label{eq:EFE}
\end{align} 

The parameters of the wormhole are $a$ and $L$, their concrete physical meaning is going to be mentioned later on. Oblate spheroidal coordinates $\{t,x,y,\varphi\}$ will be used to express the line element. They are related to those of Boyer-Lindquist $\{t,r,\theta,\varphi\}$ by the relations $Lr=x$ and $y=\cos\theta$, with $L$ being a constant parameter. We have then that

\begin{equation}
ds^2=-(dt+\Omega d\varphi)^2+e^K\Delta\left(\frac{L^2dx^2}{\Delta_1}+\frac{dy^2}{1-y^2}\right)+\Delta_1(1-y^2)d\varphi^2,
\label{eq:solBL}
\end{equation}  
with $\Delta=L^2(x^2+y^2)$, $\Delta_1=L^2(x^2+1)$,

\begin{equation}
\Omega=\frac{ax(1-y^2)}{L(x^2+y^2)}, \hspace{4mm} \text{and} \hspace{4mm} K=\frac{k}{L^4}\,\frac{\left[1-y^2\right]\left[8x^2y^2(x^2+1)-(1-y^2)(x^2+y^2)^2\right]}{(x^2+y^2)^4}.
\label{omegaK}
\end{equation}
Finally, the constant $k$ is defined by

\begin{equation}
k=\frac{a^2}{8}\left(1-\frac{4\varepsilon}{\alpha^2}\right).
\label{kmin}
\end{equation}
Interesting special cases for the coupling constant are $\alpha^2=1$, which represents a low-energy string theory, and $\alpha^2=3$, in which the Lagrangian (\ref{eq:lagrangian}) reduces to that of a 5-D Kaluza-Klein theory. In Table \ref{tablak} we show the values of the constant $k$ in the previous cases for the dilatonic and ghost fields. It also contains the value of $\alpha^2$ for which $k=0$ (only the dilatonic field is possible).

\begin{table}[h]
\caption{Real values of $k$ for some cases of $\alpha^2$ for both dilatonic and ghost scalar fields.}
\begin{center}
\begin{tabular}{ |c||c|c| } 
 \hline
   & \multicolumn{2}{|c|}{$k$} \\ \hline
 $\alpha^2$ & Dilatonic Field ($\varepsilon=1$) & Ghost Field ($\varepsilon=-1$) \\ \hline \hline
 1 & $-3a^2/8$ & $5a^2/8$ \\ \hline 
 3 & $-a^2/24$ & $7a^2/24$ \\ \hline
 4 & 0 & - \\ \hline
\end{tabular}
\end{center}
\label{tablak}
\end{table}

The scalar field $\Phi$ and the electromagnetic vector potential $A_\mu$ are given by 

\begin{equation}
\Phi=\frac{ay}{\alpha L^2(x^2+y^2)}, \hspace{4mm} A=-\frac{e^{\alpha\Phi}}{2}\left[(1-e^{-\alpha\Phi})dt+\Omega d\varphi\right],
\nonumber 
\end{equation} 
while the electromagnetic field $F_{\mu\nu}$ can be computed from $A_\mu$ as

\begin{align}
F=\frac{aLe^{\alpha\Phi}}{\Delta^2}\left[\right.&2Lxydt\wedge dx+\frac{L^2}{\Delta}\left(1-y^2\right)\left(L^2[y^4-x^4]-2ax^2y\right)dx\wedge d\varphi \nonumber\\
&+\left.L(y^2-x^2)dt\wedge dy+\frac{x}{\Delta}\left(aL^2[x^2-y^2][1-y^2]-2y\Delta\Delta_1\right)dy\wedge d\varphi\right].
\nonumber
\end{align}
It is, however, more illustrative to consider the asymptotically dominant components of this field tensor in an orthonormal frame $\{\hat{t},\hat{x},\hat{y},\hat{\varphi}\}$, namely,

\begin{equation}
F_{\hat{\mu}\hat{\nu}}=\frac{a}{L^3x^3}
\begin{bmatrix}
	0 & y & -\sqrt{1-y^2}/2 & 0 \\
	-y & 0 & 0 & -\sqrt{1-y^2}/2 \\
	\sqrt{1-y^2}/2 & 0 & 0 & -y \\
	0 & \sqrt{1-y^2}/2 & y & 0 \\ 	
\end{bmatrix}
+\mathcal{O}\left(\frac{1}{x^4}\right).
\nonumber
\end{equation}
Since asymptotically $Lx\sim l$ and $y=\cos\theta$, being $l$ and $\theta$ regular spherical coordinates in flat space-time, then the form of this electromagnetic field can be immediately identified with that of an electric and magnetic dipole. Their electric and magnetic dipole moments, $p$ and $\mu$ respectively, are $p=\mu=a/2$. Hence, in the following this metric will be referred to as the electromagnetic dipole wormhole, or electromagnetic wormhole for short\footnote{This differs from reference \cite{DelAguila} where this wormhole is claimed to be purely magnetic due to a misconception related to the use of a specific gauge to express the electromagnetic vector four-potential. Here, the orthonormal components of the electromagnetic tensor, which is gauge invariant, help us to realize that there is also a non-ignorable electric dipole contribution.}.

The calculation and the analysis of the scalar curvature invariants of this electromagnetic wormhole yield the following general form

\begin{equation}
R_X=\frac{e^{-\delta K}F(x,y)}{(x^2+y^2)^\beta},
\label{RXM}
\end{equation}
here, $\delta$ and $\beta$ are positive integers, and $F(x,y)$ is a polynomial of degree less than the degree of $(x^2+y^2)^\beta$. Particularly, $\delta=1$ for invariants of linear order in the curvature tensors, e.g., the Ricci scalar $R=R^\mu_{\ \mu}$, and $\delta=2$ for quadratic invariants, e.g., $R^{\mu\nu}R_{\mu\nu}$. Note that curvature is not well-defined at the point $x=y=0$ and its limit depends on the direction of approach. From (\ref{RXM}) it can be seen that an observer will encounter an infinite or vanishing curvature depending on the sign that $K$ takes on its path (details of this can be found in the upcoming figures \ref{fig:geoWHMP} and \ref{fig:geoWHMN}). For this reason, this kind of pathological behavior is sometimes called directional singularity.

Other relevant physical characteristics of the wormhole are:
\begin{itemize}
\item The parameter $L>0$ has units of length and is related to the size of the throat of the wormhole, while $a$ has units of angular momentum. 
\item Its mass $m$ and angular momentum $J$, which are found by using Komar integrals evaluated on two-spheres of arbitrarily large radius, are $m=0$ and $J=a$.
\item The throat of the wormhole is located at $x=0$.
\item It presents a curvature ring singularity at $x=y=0$ of radius $L$, which bounds the throat.
\item It is asymptotically flat.  
\item If $a=0$, the metric describes a flat-space time in oblate spheroidal coordinates. The relation between them and regular Cartesian coordinates $\{u_1,u_2,u_3\}$ is:

\begin{equation}
u_1=L\sqrt{(x^2+1)(1-y^2)}\cos\varphi, \hspace{4mm} u_2=L\sqrt{(x^2+1)(1-y^2)}\sin\varphi, \hspace{4mm} u_3=Lxy.
\label{u}
\end{equation}
\end{itemize} 

Unfortunately, when trying to describe the motion of freely-falling particles in this space-time, it turns out that an irreducible quadratic Killing tensor cannot be found. The existence of such a type of tensors is significantly helpful for the integrability of geodesics since they imply the presence of conserved quantities along those curves. Thus, they add a fourth constant of motion to the three that are guaranteed to exist in an axially symmetric and stationary space-time like the present wormhole. The question of whether higher order tensors of this kind may exist or not in this space-time is left opened. Nevertheless, by taking a physically meaningful limit on the metric parameters, we can obtain the desired Killing tensor. It may be worth remarking at this point that this only works as an approximation. The mentioned limit is that of a slowly rotating wormhole. Mathematically, it is expressed as a condition on the physical parameters, $a/L^2\ll1$.

After applying the slowly rotating limit, separated equations of motion for the electromagnetic wormhole can be written as (see reference \cite{DelAguila} for details),

\begin{equation}
\Delta^2\dot{x}^2=X(x), \hspace{0.5cm} \Delta^2\dot{y}^2=Y(y),
\label{xysep}
\end{equation}
where an overdot indicates differentiation with respect to an affine parameter $\lambda$, and

\begin{equation}
X(x)=\Delta_1\left[(\kappa+\mathcal{E}^2)x^2+\frac{\mathcal{K}}{L^2}\right]-\frac{2a\mathcal{EL}x}{L}+\mathcal{L}^2, \hspace{4mm} Y(y)=\left[1-y^2\right]\left[L^2(\kappa+\mathcal{E}^2)y^2-\mathcal{K}\right]-\mathcal{L}^2.
\label{Y} 
\end{equation}
In the past equations, the mentioned four constants of motion appear. Namely, $p_0=-\mathcal{E}$ and $p_3=\mathcal{L}$ are two of the conjugate momenta $p_\mu=g_{\mu\nu}\dot{x}^\nu$, both of them are conserved due to the symmetries of the wormhole. They can respectively be associated with the physical interpretation at asymptotic infinity of the energy of freely-falling test particles and their projection of angular momentum on the z-axis. The third conserved quantity is their Hamiltonian $2\mathcal{H}=\kappa=g^{\mu\nu}p_\mu p_\nu$, which takes values $\kappa=-1,0,1$ for time-like, null, and space-like geodesics, respectively. Finally, $\mathcal{K}=K^{\mu\nu}p_\mu p_\nu$ is the constant related to the existence in the slowly rotating limit of a second-rank Killing tensor $K^{\mu\nu}$. Here we will spare the details on how such a tensor is obtained using the mentioned limit, however, we must emphasize that it is based on keeping only first order terms of the quantity $a/L^2$ in the wormhole metric. So, for instance, we can approximate $$e^K\approx1+\mathcal{O}\left(\frac{a^2}{L^4}\right).$$ By looking at the explicit expression for $K$ in (\ref{omegaK}), it is easy to realize that, as we get closer to the ring singularity, the approximation begins to disagree with respect to the original function. Thus, the slowly rotating limit is no longer valid for regions very close to the ring singularity.

Despite the restricted validity of the slowly rotating limit, it was argued in \cite{DelAguila} that geodesics in general were repelled when trying to reach the ring singularity at $x=y=0$. The analysis therein used the fact that the polynomials $X(x)$ and $Y(y)$ must be equal to strictly positive quantities, as can be seen from the equations in (\ref{xysep}), and also that $X(0)=-Y(0)=\mathcal{K}+\mathcal{L}^2$. This statement applies to geodesics for which $\mathcal{K}\neq-\mathcal{L}^2$. In the special case of geodesic congruences with $\mathcal{K}=-\mathcal{L}^2$, though, one cannot so easily guarantee the same property.\footnote{As a clarifying note we point out that, even though at exactly the singular region the slowly rotating limit breaks down, its valid domain is close enough to it such as to obtain a negative value in either of the $X(x)$ or $Y(y)$ polynomials. This leads to a repulsive effect on the geodesics with $\mathcal{K}\neq-\mathcal{L}^2$ owing to an infinite potential barrier around the singularity. Unfortunately, this does not happen for values $\mathcal{K}=-\mathcal{L}^2$.} These curves were not studied with detail in the cited reference and hence, during the rest of this paper, we will focus strictly on their possible incompleteness as a consequence of the singularity. 

\section{III. Beyond the Validity of the Slowly Rotating Limit}

In what follows we wish to describe the behavior of geodesics in regions where the slowly rotating limit breaks down. The inconvenience with the combination of conserved quantities described by the condition $\mathcal{K}+\mathcal{L}^2=0$ is that, according to the equations of motion (\ref{xysep}), the negative potential barrier disappears when approaching the singularity (when $x,y\rightarrow0$). Consequently, those curves can continue its trajectory into the ill-defined region of the space-time. Eventually the slowly rotating limit will break down, and the separable equations (\ref{xysep}) will no longer give reliable information about the motion of test particles. 

The Hamiltonian, $2\mathcal{H}=\kappa$, keeps holding without the need of approximations. It can be rearranged in the following convenient form,

\begin{equation}
e^K\Delta\left(\frac{L^2\dot{x}^2}{\Delta_1}+\frac{\dot{y}^2}{1-y^2}\right)=\kappa+\mathcal{E}^2-\frac{(\Omega\mathcal{E}+\mathcal{L})^2}{\Delta_1(1-y^2)}.
\label{HSFWH}
\end{equation}
This expression, nonetheless, is not enough to determine the path geodesics will follow in a general case. The only resource left then is to consider the geodesic equation, which is of course valid everywhere but extremely difficult (if not impossible) to study in full detail analytically. However, some special and interesting cases can be considered with the help of (\ref{HSFWH}). 

\subsection{III.A. Motion in the equatorial plane outside the throat}

In order for geodesics to be constrained to the plane $y=0$, we also require that $\dot{y}$ and $\ddot{y}$ both vanish. The first condition can only be achieved in $y=0$ if $\mathcal{K}+\mathcal{L}^2=0$. Then, when setting $y=\dot{y}=0$, the $\ddot{y}$ component of the geodesic equation (not shown here) vanishes too and this type of motion is indeed possible without imposing further restrictions. 

We are now interested in the $\ddot{x}$ component of the geodesic equation, but instead of looking into its full expression, we can use (\ref{HSFWH}) with $y=\dot{y}=0$ to write $\dot{x}$ as

\begin{equation}
e^{K_x}\frac{L^4x^2\dot{x}^2}{\Delta_1}=\kappa+\mathcal{E}^2-\frac{1}{\Delta_1}\left(\frac{a\mathcal{E}}{Lx}+\mathcal{L}\right)^2,
\label{Hx}
\end{equation}
with $K_x=\left.K\right|_{y=0}$. Note that the left hand side of the previous equation is positive. However, as $x\rightarrow0$, the term in round brackets of the right hand side dominates over the others. This is a negative term, thus implying a contradiction since clearly, a positive quantity cannot be equal to a negative one. By this argument, it can be established that there are no solutions of (\ref{Hx}) that reach the ring singularity. This fact was already mentioned in reference \cite{DelAguila}.

\subsection{III.B. Motion within the throat}

Geodesics that are constrained to the throat of the wormhole satisfy $x=\dot{x}=\ddot{x}=0$. Again, if $\mathcal{K}+\mathcal{L}^2=0$, then $\dot{x}=0$ in the throat. From the $\ddot{x}$ component of the geodesic equation with vanishing $x$ and $\dot{x}$ we have that,  

\begin{equation}
\ddot{x}-e^{-K_y}\frac{a\mathcal{E}\mathcal{L}}{L^5y^4}=0.
\nonumber
\end{equation}
Here, $K_y=\left.K\right|_{x=0}$. It can be seen that this type of motion is then only possible if any of the conserved quantities $\mathcal{E}$ or $\mathcal{L}$ are zero. Applying the same analysis as before, we utilize (\ref{HSFWH}) now with $x=\dot{x}=0$, obtaining thus,

\begin{equation}
\dot{y}^2=\frac{e^{-K_y}}{L^4y^2}\left[L^2(1-y^2)(\mathcal{E}^2+\kappa)-\mathcal{L}^2\right].
\label{Hy}
\end{equation}
In what follows we will consider that $\mathcal{E}\neq0$ and $\mathcal{L}=0$, due to the fact that this corresponds to a familiar and physically realistic case, in contrast to the other possibility of a test particle with vanishing energy. With the help of equation (\ref{Hy}) it can now be determined if geodesics that lie in the throat can get arbitrarily close to the singularity of the space-time.

For a wormhole with $k\geq0$, $\dot{y}^2\rightarrow\infty$ as $y\rightarrow0$. Hence, these geodesics can infinitesimally approach the singularity, possibly becoming incomplete. In fact, equation (\ref{Hy}) can be easily integrated if $k=0$ and for the particular case of null geodesics with vanishing angular momentum. The solution simply reads 

\begin{equation}
y=\pm\sqrt{1-\lambda'^2},
\label{solincomp}
\end{equation} 
where the affine parameter $\lambda'$ was rescaled though a linear transformation of the original parameter $\lambda$ for convenience. 
When $\lambda'=1$ the singularity is clearly reached by these curves, which are only defined for $\lambda'\in[-1,1]$.

On the other hand, geodesics in wormholes with $k<0$ behave in the opposite way, as $y\rightarrow0$ we have that $\dot{y}^2\rightarrow0$. In fact, $\ddot{y}\rightarrow0$ as $y\rightarrow0$ too. This can be seen from the $\ddot{y}$ component of the geodesic equation of these particular curves:

\begin{equation}
\ddot{y}+\frac{\left[L^4y^4+2k(1-y^2)^2\right]\dot{y}^2}{L^4(1-y^2)y^5}=0.
\nonumber
\end{equation} 

The combination of these kinematic conditions indicates that geodesics constrained to the throat and that start their path at some value $y=y_0$ will slow down when approaching the singularity. As they go closer, their coordinate velocity $\dot{y}$ will become smaller, almost completely decreasing to zero. As a result, they will never reach the ring singularity in a finite amount of their affine parameter, in other words, an infinite affine parameter is needed so that they can meet the singularity.

This past conclusion can also be obtained from the following approximate solution of equation (\ref{Hy}) for $y\ll1$,

\begin{equation}
\lambda'=\pm\int ye^{-k/2L^4y^4}dy,
\label{solincomp2}
\end{equation}
here the original affine parameter $\lambda$ was properly rescaled again for simplicity. It is readily seen that if $k\geq0$, the integrand vanishes as $y\rightarrow0$, leading to a finite affine parameter at which the singularity is met. On the contrary, with $k<0$ the integrand becomes infinite at $y=0$ and consequently, so does $\lambda'$ when trying to reach the singular region. It should be remarked that, of the interesting cases for the coupling constant $\alpha$ shown in table \ref{tablak}, those that have a negative $k$ correspond to a dilatonic field, whereas those with positive $k$ represent a ghost field. Also from equation (\ref{kmin}) and for a ghost field ($\varepsilon=-1$) we have that $k>0$. 

The two previous simple types of motion have already given valuable information. The most important is that equatorial geodesics are not in contact with the ring singularity. In contrast, those that lie within the throat are able to encounter it in a finite affine parameter if $k\geq0$. Indeed, equations (\ref{solincomp}) and (\ref{solincomp2}) indicate that this is the case for geodesics with zero angular momentum. This rules out said wormholes as space-times possessing a curvature singularity without geodesics touching it. In fact, only the electromagnetic dipole wormhole metric with negative $k$ stands now as a possible candidate of such a space-time.  

\subsection{III.C. Numerical analysis of general geodesics}
\label{ss:geonum}

There are of course more general geodesics other than those constrained to the equatorial plane or the throat. However, due to the complicated expressions of the geodesic equation for these wormholes, it is not possible to analytically solve these curves. Since according to the previous results, there exists the possibility that geodesics can be found arbitrarily close to the ring singularity by traveling near the disc bounded by it, we need to determine if there are general geodesics (others than those of constant $x=0$) that follow this path. The only resource left is to study them numerically with the inherent and unfortunate restrictions of these methods.

\begin{figure}[h]
	\centering
		\includegraphics[scale=0.28]{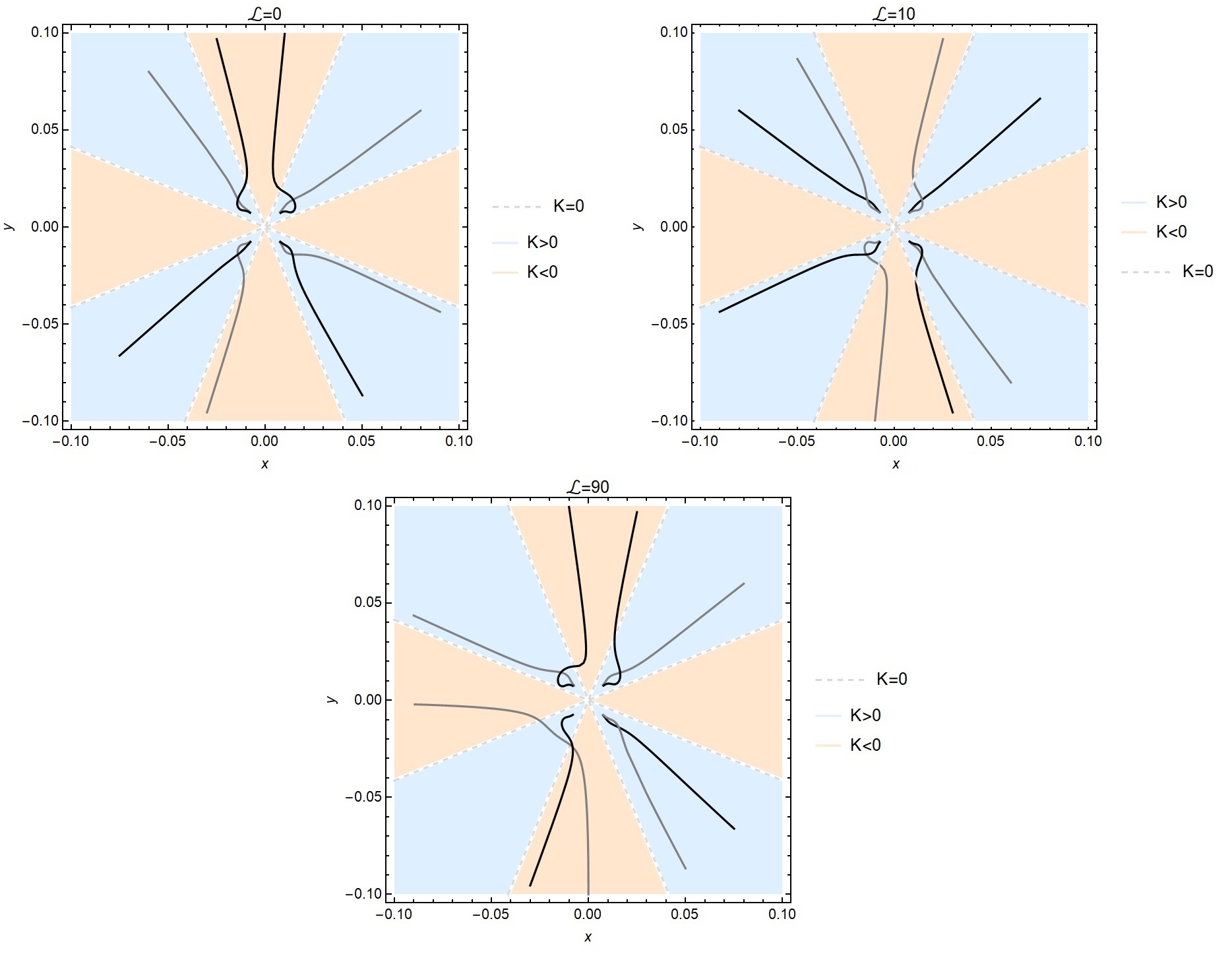}
		\caption{Null geodesics in the electromagnetic wormhole with $k=7a^2/24$ (a ghost field) for three values of angular momentum $\mathcal{L}$. The numerical values of the space-time parameters and constants of motions are $a=0.1$, $L=10$ and $\mathcal{E}=10$. Different colors are used to distinguish between intersecting curves.}
			\label{fig:geoWHMP}
\end{figure}

The procedure we shall follow is now described. We are interested in geodesics for which $\mathcal{K}+\mathcal{L}^2=0$ in the slowly rotating limit. Therefore, initial conditions for the coordinates $x$ and $y$ are chosen in a region where this limit is valid. In turn, for given values of the constants of motion $\mathcal{E}$, $\mathcal{L}$ and $\kappa=0,-1$, as well as the corresponding space-time parameters, the initial position $x_0$ and $y_0$ will fix the initial velocities $\dot{x}_0$ and $\dot{y}_0$ according to equations (\ref{xysep}). Finally, as is evident, our principal objective is to examine geodesics that advance toward the singularity as their affine parameter increases. With this set of initial data and conditions the numerical calculation of geodesics is performed. Several results are shown in the following figures, presenting the curves in the $x$-$y$ plane of coordinates (the singularity is described by the origin). 

As discussed previously in the analysis of the two simple types of motion, we can expect different behaviors in this wormhole depending on the sign of the constant $k$ of equation (\ref{kmin}). We thus start with the case of a positive $k$ (figure \ref{fig:geoWHMP}). An interesting property that seems to dominate the path taken by geodesics is whether $K$ is positive or negative in the $e^{-K}$ factor appearing, for example, in equations (\ref{RXM}) and (\ref{Hy}). Curves whose initial position is in the $K>0$ area tend to stay within it, and curves whose initial position is in the $K<0$ area tend to cross their starting region into the first area. Furthermore there are four points in the $x$-$y$ plane, one for each $K>0$ zone, at which all curves seemingly converge to. The only found exceptions to this behavior were already described, i.e., motion constrained to $y=0$ or $x=0$ (not shown in figure \ref{fig:geoWHMP}). It must be mentioned that although it appears that geodesics reach one of these supposed convergence points and then stop there, it is not quite exactly what happens. Instead, their velocities and accelerations become increasingly small as they advance, most likely as a consequence of the $e^{-K}$ factor severely decreasing as well. This is not at all unfamiliar since we found earlier that some geodesics in the throat of the wormhole with $k<0$ exhibit this type of behavior. Other than geodesics constrained to the throat, which can exist arbitrarily close to the singularity, no other curves were found that could touch it in a finite amount of affine parameter. Due to the fact that geodesics in the throat become incomplete, we should definitely discard this kind of wormhole as being regular. One must keep in mind, though, that the problematic geodesics are bounded by the ring, they do not escape to infinity. In this sense, one can think of this space-time as not so badly behaved for distant observers. It is worth remarking that $k$ is positive for any wormhole coupled to a ghost field and therefore, all such space-times are geodesically incomplete.

\begin{figure}[h]
	\centering
		\includegraphics[scale=0.28]{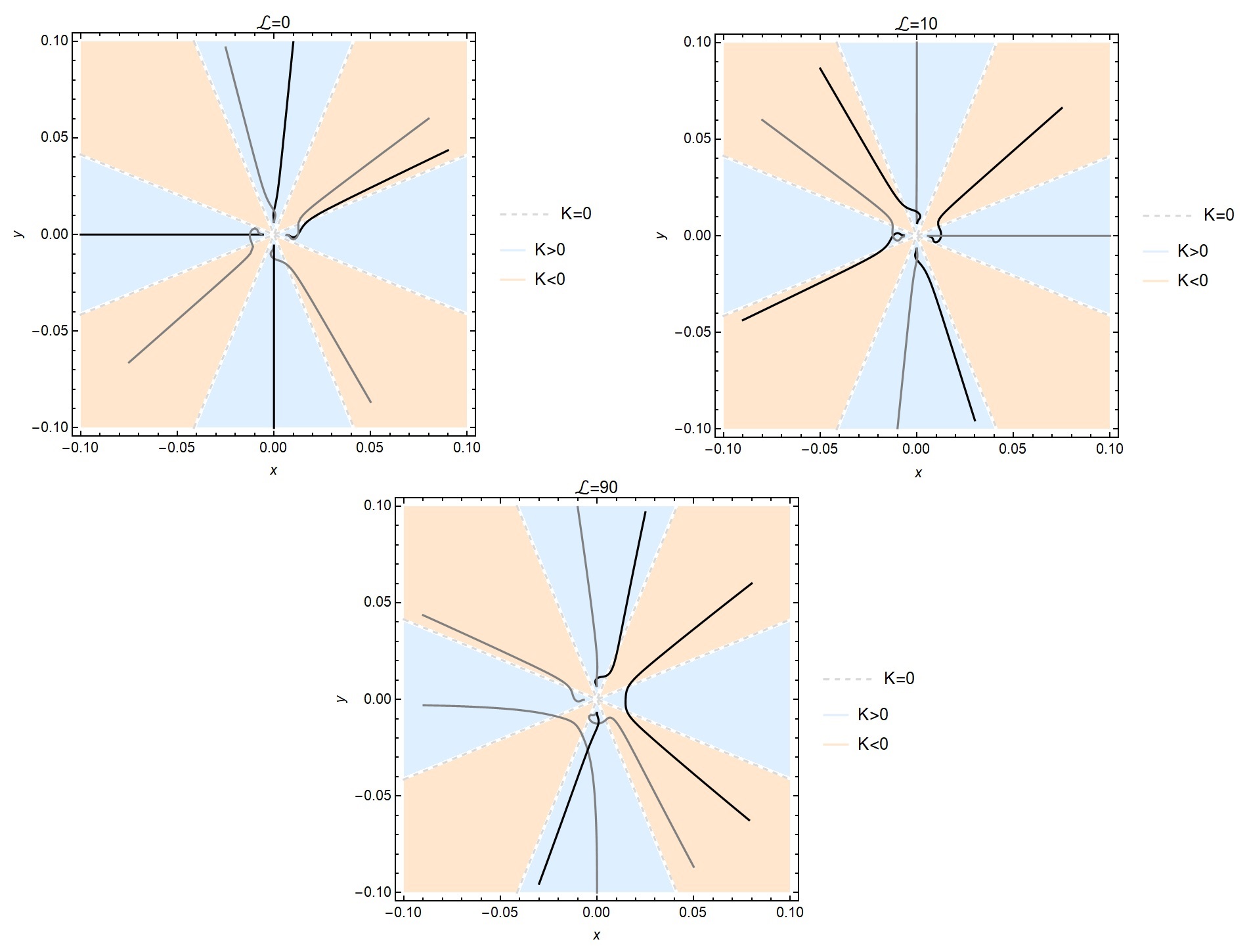}
		\caption{Null geodesics in the electromagnetic wormhole with $k=-a^2/24$ (a dilatonic field) for three values of angular momentum $\mathcal{L}$. The numerical values of the space-time parameters and constants of motions are the same as those of previous figures. Different colors are used to distinguish between intersecting curves.}
			\label{fig:geoWHMN}
\end{figure}

There are some similarities between the electromagnetic dipole wormhole with $k>0$ and that with $k<0$ (figure \ref{fig:geoWHMN}). The behavior of the curves is naturally also heavily influenced by the $e^{-K}$ factor, but as the sign of the constant $k$ has been changed, so have the areas of positive and negative $K$. We also observe here apparent convergence points that were rotated so that they are now located in the $K>0$ region too. From these points forward, and in direction to the singularity, the curves seem to greatly slow down as well. The difference in this wormhole is that the inversion of the $K$ regions with respect to the $k>0$ case allows for the $x$ and $y$ axis to be found within a positive $K$ area. This leads to the past result in which geodesics constrained to $x=0$ take an infinite amount of affine parameter to meet the singular ring. Therefore it could be said that, in a strict sense, causal geodesics do not encounter the curvature singularity. Furthermore, none of this type of curves were found to be incomplete either. Hence, the electromagnetic dipole wormhole with a negative constant $k$ remains as a candidate of a space-time with complete curves despite the presence of curvature singularities. Only wormholes coupled to dilatonic fields can feature this property. Nevertheless, since not all values of the coupling constant $\alpha^2$ lead to $k<0$ in (\ref{kmin}), not all of the dilatonic wormholes ($\varepsilon=1$) will be geodesically complete. For reference, the coupling constants corresponding to a low energy-string theory and to Kaluza-Klein theory have a negative $k$ parameter (see table \ref{tablak}).

This specific manner of achieving completeness in the analyzed geodesics leads to peculiar conclusions that resemble properties of the so-called ``bag of gold'' singularities introduced by Wheeler in \cite{lehouches}. Namely, the counter-intuitive notion of having an infinite volume bounded by a finite superficial area. To see this, consider a surface $S_\epsilon$ of $t$ and $x=\epsilon$ constant. It is evident that $S_\epsilon$ has a finite superficial area. Now take a space-like geodesic $\zeta$ with $t$, $\varphi$ and $x=0$ constant ($\mathcal{E}=\mathcal{L}=0$ and $\kappa=1$) in (\ref{Hy}). The approximate solution in (\ref{solincomp2}) is equally valid for this case, hence, the geodesic length of this curve becomes infinite when approaching the singularity. The geodesic $\zeta$ lies inside the volume $V_\epsilon$ bounded by $S_\epsilon$ and thus, it would be reasonable to compute this volume using the geodesic length of said curve. However, by doing this, one ends up with an infinite volume bounded by a finite superficial area. This unusual property is a direct consequence of having geodesics directed toward the singularity, but never completely reaching it. 

As mentioned in section I, the atypical behavior displayed by the previously described causal geodesics is not completely new in the literature. Another work worth mentioning that contemplates this scenario (besides \cite{sussman}) can be found in \cite{gcbh}. The space-times of interest there, however, are different from those of this paper in the sense that possible geometries of geodesically complete and spherically symmetric black holes are discussed. It is also assumed that curvature singularities are generally absent from the metric. Despite this, the comprehensive analysis done therein goes as far as to include a case in which curvature singularities are indeed present and outgoing null geodesics are separated from them by an infinite affine length. The example presented in this paper replicates that feature but for general causal geodesics of an axially symmetric wormhole. It should be remarked that other assumptions made for the geometry of the space-times from \cite{gcbh} include global hyperbolicity, a property that is yet to be analyzed for this type of wormhole.

Some final comments are now outlined. These figures only show null geodesics, the reason for this is that there was no significant qualitative difference between said geodesics and their time-like counterparts. Lastly, the results described here are numerous, in order to concisely present them, table \ref{tab:WH} summarizes their most important aspects. 

\begin{table}[h] 
\caption{The properties of geodesics that closely approach the ring singularity $\sigma$ of the electromagnetic dipole wormhole.}
\begin{center}
\begin{tabular}{ |c||c|c|c| }
 \hline
   & \multicolumn{2}{|c|}{Geodesics that encounter $\sigma$} & \\ \hline
 Wormhole & Conditions & Curvature & Regularity  \\ \hline \hline
  & Constrained to $x=0$ & & Singular (singularity \\ 
$k\geq0$ & (finite $\lambda$ required, & Unbounded & visible to observers \\ 
& incomplete) & & in the throat) \\ \hline
 $k<0$ & Infinite $\lambda$ required & Vanishing & Complete causal \\ 
 & to reach $\sigma$ & & geodesics \\ \hline
\end{tabular}
\end{center}
\label{tab:WH}
\end{table}

\section{IV. Five-Dimensional Interpretation of the Singularity}

In this section we will use the five-dimensional Kaluza-Klein theory in order to probe, in the context of a higher dimensional space, the geometry of a neighborhood of the singularity. For this purpose, embedding diagrams shall be utilized, thus allowing us to visualize slices of the five-dimensional metric corresponding to the wormhole studied so far.

In Kaluza-Klein theory an extra dimension is added to the typical four dimensions of General Relativity. This is done as a means to naturally incorporate electromagnetism into the geometry of the space-time, very much as the gravitational force arises purely from the curvature of a four-dimensional manifold in standard gravity. In this sense, the theory could be viewed as an extension of General Relativity that unifies gravity with electromagnetism.

The geometric properties and the consequences of an additional dimension may be physically hard to grasp. However, in effective terms it is interpreted as being compactified and hence, imperceptible to a macroscopic observer which would experience space-time consisting only of four dimensions. In fact, Klein suggested that the geometry of the extra dimension is of the form of a circle $S^1$ with a small radius. We will follow this idea and therefore, shall consider the fifth dimension as periodic.

The five-dimensional metric $g_{AB}$ is given by

\begin{equation}
ds^2=g_{AB}dx^A dx^B=I^{-1}g_{\mu\nu}dx^\mu dx^\nu+I^2(dx^5+A_\mu dx^\mu)(dx^5+A_\nu dx^\nu),
\label{ds25D}
\end{equation}
with $I=e^{\alpha\Phi/3}$. The parameter $\alpha$ is the coupling constant discussed earlier. Also $\Phi$ and $A_\mu$ are, respectively, the scalar field and the electromagnetic four-potential (if any) coupled to the original four-dimensional solution. In this 5-D analysis Latin indices are used which take values $A,B=1,2,\ldots,5$. Then, the Einstein-Hilbert action in five dimensions can be written as,

\begin{equation}
S^{(5)}=\frac{1}{16\pi G^{(5)}}\int d^5x\sqrt{-g^{(5)}}R^{(5)},
\nonumber
\end{equation}
where a $(5)$ superscript denotes quantities of the five-dimensional space-time. Electromagnetic interactions in four dimensions arise from vacuum solutions obtained from this action if they are projected into the standard four-dimensional space-time. When inserting the metric of equation (\ref{ds25D}) in this action, and after integrating over the periodic fifth coordinate $x^5$, the action in four dimensions resulting from Lagrangian (\ref{eq:lagrangian}) with $\varepsilon=1$ and $\alpha=\sqrt3$ is obtained. Note hence that the studied wormholes coupled to phantom fields ($\varepsilon=-1$) cannot be described by this theory. See \cite{KK} and references therein for a more in-depth review of the subject.

We follow \cite{5D} and start by taking slices of constant $t$ and $\varphi$ in metric (\ref{ds25D}), reducing it to a three-dimensional subspace. Thus,\footnote{In some of the following equations, the tensor product $\otimes$ will be explicitly written just to avoid any sort of possible confusion caused by exponents in the infinitesimal coordinate displacements $dx^A$.}

\begin{equation}
ds^2_{3D}=I^{-1}g_{ij}dx^i\otimes dx^j+I^2dx^5\otimes dx^5, \hspace{4mm} (i,j=2,3), 
\nonumber 
\end{equation}
only if $A=A_tdt+A_\varphi d\varphi$, which is the case for the stationary and axially symmetric wormhole of interest. The resulting subspace is going to be even further reduced to two dimensions in order to properly embed it in a three-dimensional Euclidean space. The process shall be done separately for constant $x^2=x_0^2$ and constant $x^3=x_0^3$, i.e., for constant $x=x_0$ and $y=y_0$ in the oblate spheroidal coordinate system $\{t,x,y,\varphi\}$ used in the paper. This yields,

\begin{equation}
ds^2_x=I_2^{-1}g_{xx}dx\otimes dx+I_2^2dx^5\otimes dx^5, \hspace{4mm} ds^2_y=I_3^{-1}g_{yy}dy\otimes dy+I_3^2dx^5\otimes dx^5,
\label{ds2i}
\end{equation}
where $I_2=I(x,y_0)$ and $I_3=I(x_0,y)$. Expressed more compactly we have that, $$ds^2_i=I_i^{-1}g_{ii}dx^i\otimes dx^i+I_i^2dx^5\otimes dx^5, \hspace{2mm} \text{for fixed values } i=2,3 \text{ (no sum over repeated indices).}$$ In equation (\ref{ds2i}) we are already focusing on the form of metric (\ref{eq:solBL}), nevertheless, this procedure is possible for any stationary and axially symmetric space-time that admits an adapted coordinate system (one in which the only off-diagonal metric component is $g_{t\varphi}$). To embed these two subspaces in three dimensions we follow the typical scheme. First we take the line element of flat space-time in cylindrical coordinates, $ds^2_{cyl}=dz^2+d\rho^2+\rho^2d\phi^2,$ and consider $z(\rho)$ as a profile function that describes the embedded geometry. Alternatively, one can instead make $z(x^i)$ and $\rho(x^i)$ ($i=2,3$) so that each coordinate value $x^i$ defines a point in the $z$-$\rho$ plane, hence obtaining the desired profile. So, $ds^2_{cyl}$ can be rewritten as

\begin{equation}
ds^2_{cyl}=\left[\left(\frac{dz}{dx^i}\right)^2+\left(\frac{d\rho}{dx^i}\right)^2\right]dx^i\otimes dx^i+\rho^2(x_i)d\phi\otimes d\phi.
\label{ds2cyl}
\end{equation}  
Since the fifth coordinate $x^5$ is assumed to be periodic, it can be associated with the azimuthal angle $\phi$ of the previous flat space. Comparing (\ref{ds2cyl}) with (\ref{ds2i}) we obtain,

\begin{equation}
\left(\frac{dz}{dx^i}\right)^2+\left(\frac{d\rho}{dx^i}\right)^2=\frac{g_{ii}}{I_i}, \hspace{4mm} \rho^2(x^i)=I_i^2.
\label{dzdxi}
\end{equation} 

This defines a first order differential equation for $z(x^i)$ that later on will be solved numerically for each value $i=2,3$. Before that, interesting properties of the embedded profile can be found from $\rho(x^i)=I_i$, which is basically a relation for the radius of the fifth dimension. According to the ongoing analysis, this radius will depend on whether we are interested in slices of constant $y=y_0$ or constant $x=x_0$, it is respectively given by $$I_2=e^{ay_0/3L^2(x^2+y_0^2)}, \hspace{4mm} \text{or} \hspace{4mm} I_3=e^{ay/3L^2(x_0^2+y^2)}.$$ From these explicit expressions it is easy to make the following observations. If $y_0>0$, the radius $I_2$ has an absolute maximum at $x=0$, i.e., at the throat of the wormhole. This maximum becomes a minimum if $y_0<0$. When $x_0\neq0$, the radius $I_3$ has an absolute maximum at $y=x_0$ and an absolute minimum at $y=-x_0$. Since $y\in[-1,1]$, the maximum and minimum are only relevant when $-1\leq x_0<0$ or $0<x_0\leq1$. The general radius $I(x,y)$ is severely discontinuous at $x=y=0$, this can be seen from the fact that $I(x,0)=\left.I_2\right|_{y_0=0}=1$, and when inspecting $I(0,y)=\left.I_3\right|_{x_0=0}$ with the following limits: $$\lim_{y\rightarrow0^+}\left.I_3\right|_{x_0=0}=\infty, \hspace{4mm} \lim_{y\rightarrow0^-}\left.I_3\right|_{x_0=0}=0,$$ where we have assumed that $a>0$. Such problematic behavior, of course, is naturally expected due to the presence of the curvature singularity. In this case, we can see that it has a deep impact on the geometrical characteristics of the fifth dimension when interpreted as periodic and visualized in an Euclidean three-dimensional space.

\begin{figure}[h]
	\centering
		\includegraphics[scale=0.3]{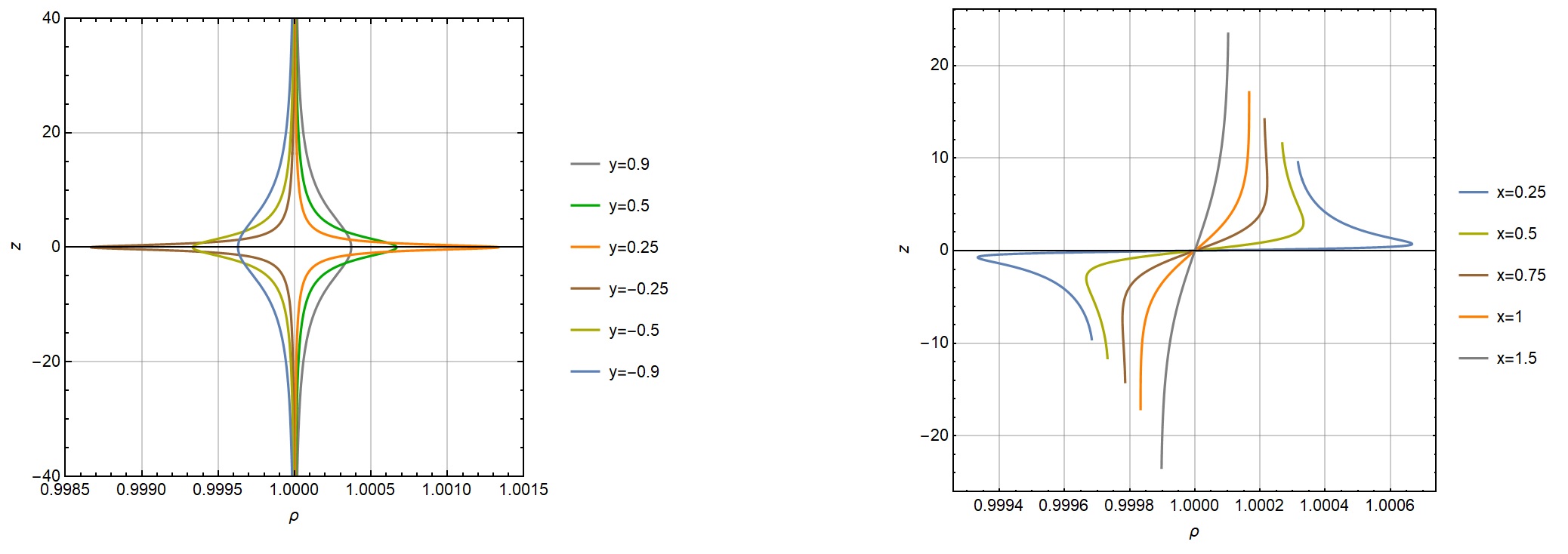}
		\caption{Embedding profiles for slices of different values $y_0\neq0$ (left panel) and $x_0\neq0$ (right panel). The domain of numerical integration is respectively $-5\leq x\leq5$ and $-0.99\leq y\leq0.99$. The resulting 2-manifolds are obtained by rotating each profile about the $\rho$ axis. The numerical values of the space-time parameters are the same as those of previous figures.}
			\label{fig:zreg}
\end{figure}

\begin{figure}[h]
	\centering
		\includegraphics[scale=0.28]{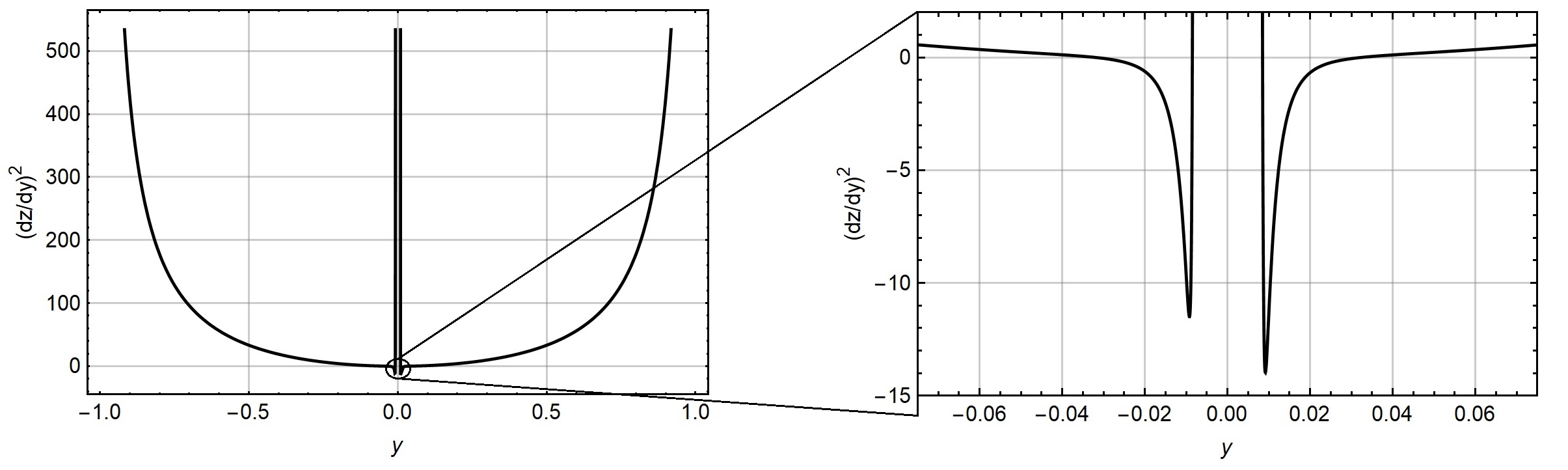}
		\caption{The function $(dz/dy)^2$ given by equation (\ref{dzdysing}). Observe that it becomes negative for some small values of $\left|y\right|$, but then it diverges to positive infinity as $y\rightarrow0$. In this case its roots are approximately at $y_{\pm1}=\pm0.0325$ and at $y_{\pm2}=\pm0.008$. The numerical values of the space-time parameters are the same as those of previous figures.}
			\label{fig:dzdy2}
\end{figure}

\begin{figure}[h]
	\centering
		\includegraphics[scale=0.35]{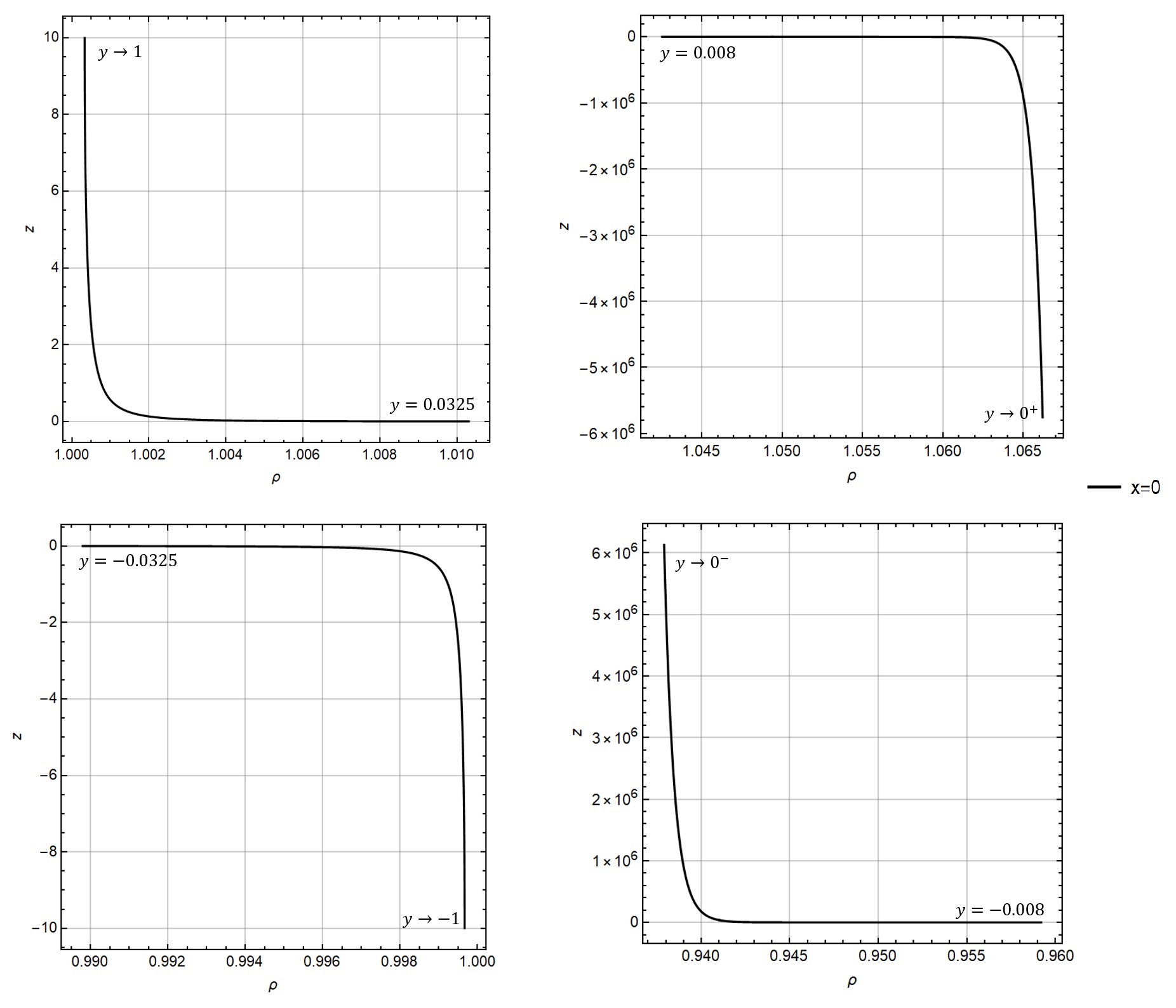}
		\caption{Embedding profiles for slices of constant $x_0=0$. The domains of numerical integration are $0.0325\leq y\leq0.99$ (top left panel), $0.005\leq y\leq0.008$ (top right panel), $-0.99\leq y\leq-0.0325$ (bottom left panel), $-0.008\leq y\leq-0.005$ (bottom right panel). The resulting 2-manifolds are obtained by rotating each profile about the $z$ axis. The values of the space-time parameters are the same as those of previous figures.}
			\label{fig:zsing}
\end{figure}

\begin{figure}[h]
	\centering
		\includegraphics[scale=0.33]{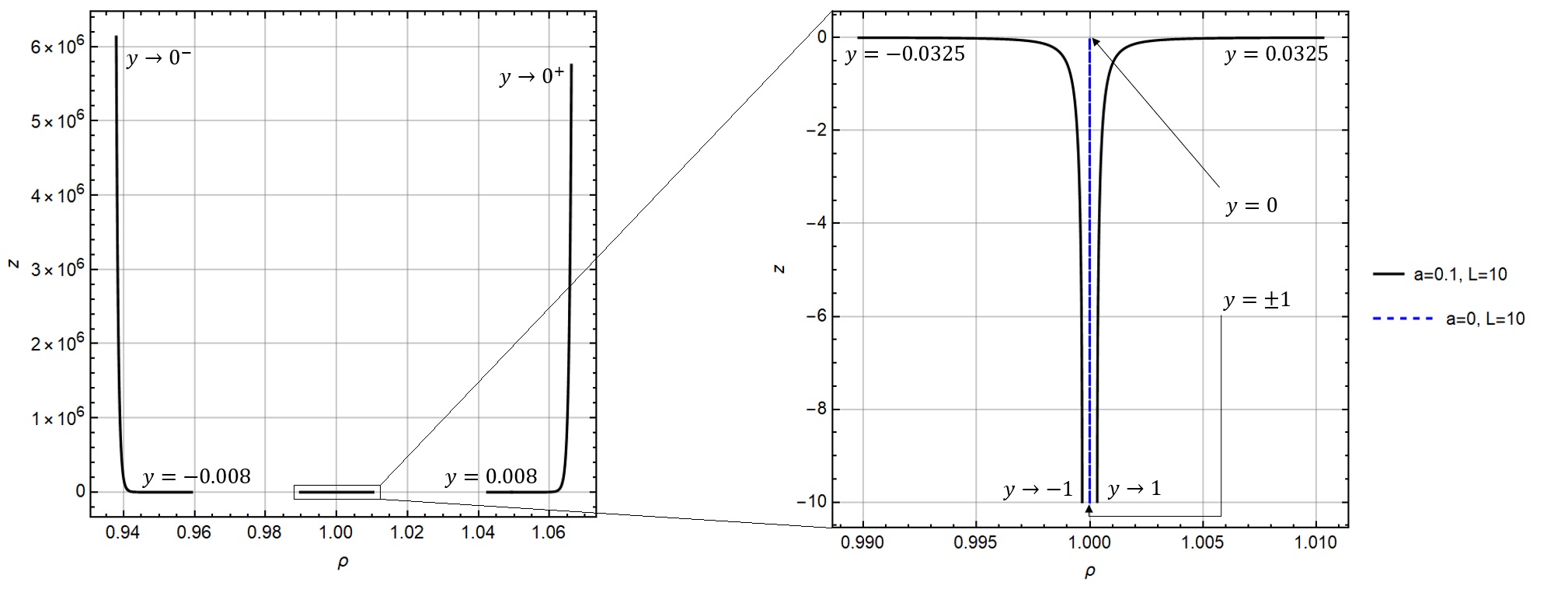}
		\caption{The embedding profiles shown in figure \ref{fig:zsing} reorganized in one general plot. The curve corresponding to values between $y=\pm0.008$ and $y=\pm0.0325$ cannot be visualized in Euclidean three-space. The profile generated by a flat space-time ($a=0$) (dashed blue line) is also included for comparative purposes.}
			\label{fig:zsing2}
\end{figure}

We now turn to the profiles $z(\rho)$ that describe the embedded geometry when considering slices of either $y$ or $x$ constant in the space-time. They are respectively obtained numerically by solving the equations in (\ref{dzdxi}) for $i=2$ and $i=3$. The profiles, which are shown in figures \ref{fig:zreg} through \ref{fig:zsing2}, are then parametrized by $\rho(x^i)$ and $z(x^i)$. For the sake of completeness, and to illustrate some of the discussed features for the radii $I_{2,3}$, we begin in figure \ref{fig:zreg} by examining the slices of $x_0\neq0$ and $y_0\neq0$, all of them being regular profiles as expected. Initial conditions were chosen so that $z(x=0)=0$ and $z(y=0)=0$ for every graphic. The upper half of the planes correspond to positive $x$ (left panel) and positive $y$ (right panel\footnote{For the domain of the $y$ variable, the points $y=\pm1$ could not be included due to the nature of the coordinates used here. Indeed, they represent the coordinate singularity of the rotation axis, very much like that appearing when using spherical coordinates.}), while the lower half to their negative counterparts. The mentioned properties for the maximum and minimum radii can be readily appreciated. Furthermore, in both cases the value of the maximum increases as the profiles draw closer to the singularity and, on the contrary, the minimum decreases approaching it. This will be consistent with the later analysis for the neighboring regions of the curvature pathology. Lastly, since the differential equation that yields the profiles for $x=x_0$ constant is symmetric for the $x$ coordinate, then the profiles obtained by taking $x=-x_0$ are the same as those with $x=x_0$ constant.

Next, we attempt to study the embedded geometry for regions close to the singularity. To do so, let us set $x_0=0$ in (\ref{dzdxi}) for $i=3$, thus probing the geometry at the throat of the wormhole. Explicitly we have that 

\begin{equation}
\left(\frac{dz}{dy}\right)^2=\frac{L^2y^2e^{K_y}}{I_3(1-y^2)}-\frac{a^2I_3^2}{9L^4y^4}, \hspace{4mm} \text{with } I_3=I(0,y)=e^{a/3L^2y} \hspace{2mm} \text{and } K_y=\left.K\right|_{x=0}=\frac{a^2(1-y^2)^2}{24L^4y^4}.
\label{dzdysing}
\end{equation}
It is important to notice some aspects of the previous equation. First, due to the squared derivative, it will generally possess two families of solutions, one corresponding to the positive branch and the other to the negative branch. Second, it is singular at $y=0$ and thus, solutions for the initial value problem $z(0)=z_0$ are not guaranteed to exist. The following limit does exist and a careful evaluation of it yields, $$\lim_{y\rightarrow0}\left(\frac{dz}{dy}\right)^2=\infty.$$ Third, the right hand side of equation (\ref{dzdysing}) can be negative for some $y\in[-1,1]$, figure \ref{fig:dzdy2} serves as an example of this, and hence no real solutions exist in such regions. This portion of the geometry cannot be visualized in three-dimensional Euclidean space. Unfortunately, the relation in (\ref{dzdysing}) is non-linear for $y$ and its roots cannot be found analytically. 

For reference, it will be useful to first study the $a=0$ case in expression (\ref{dzdysing}), i.e., a flat space-time. The differential equation then reduces simply to

\begin{equation}
\left(\frac{dz}{dy}\right)^2=\frac{L^2y^2}{1-y^2}, \hspace{4mm} \text{and } I_3=I(0,y)=1,
\label{dzdyf}
\end{equation}
which can be easily solved, obtaining thus $z(y)=\pm L(\sqrt{1-y^2}-1)$. Initial conditions were chosen so that $z(0)=0$. Considering $z(y)$, along with $\rho(y)=1$, the curve parametrized by $y\in[-1,1]$ in the $\rho$-$z$ is a compact, vertical line that joins the points $(1,\pm L)$ and $(1,0)$. This defined profile then needs to be rotated about the symmetry axis (the $z$ axis), therefore describing a cylinder of radius equal to one and of height $L$. It is easy to understand the meaning of this result. In four-dimensional space-time, $x=0$ represents a disc of radius $L$ that, by fixing the azimuthal angle $\varphi$ to a constant, becomes a line of length $L$. When introducing a periodic fifth dimension the line generates a closed cylinder of the same height. In this case, the radius of the new dimension is normalized to unity. Note also that $z(y)=z(-y)$, this is consistent with the fact that in $x=0$, $y$ and $-y$ describe the same set of points on the disc (see the equations shown in (\ref{u})).

Being aware of these characteristics for $(dz/dy)^2$, we are forced to study separately the regions in which the geometry can be embedded in a three-dimensional space, i.e., those with $(dz/dy)^2\geq0$. According to figure \ref{fig:dzdy2} there are four of them. Our interest will focus mainly on those closer to the singularity. The procedure we shall follow is to fix the initial conditions $z(y_{\pm1})=0$ and $z(y_{\pm2})=0$, where $y_{\pm1}$ and $y_{\pm2}$ are the approximate roots of $(dz/dy)^2$ mentioned in the past figure. Since the differential equation is singular when $y=0$ and $y=\pm1$, this process will create four disjoint profiles of constant $x_0=0$ that span all $y$ such that $(dz/dy)^2\geq0$, their domains being $y\in(\pm1,y_{\pm1}]$ and $y\in[y_{\pm2},0)$. These profiles are presented in figure \ref{fig:zsing}, in which the positive branch of the solutions is used in all cases.

From the figure of space-time slices corresponding to $x_0=0$ we can see that, as expected, the effects of the fifth dimension are stronger near the singularity ($\left|y\right|\ll1$). This is seen respectively in the increase or decrease of the radius when approaching it from positive or negative $y$ values. Recall that applying the same analysis to a flat space-time, which can be recovered by fixing $a=0$, the radius of the extra dimension is constant, it reduces to unity. Thus, the change of this geometrical property can be interpreted as the fifth dimension finally becoming non-trivial, or non-negligible, in the neighborhood of the singular region. In these embeddings within Euclidean three-space, the $z$ coordinate diverges to $\pm\infty$ as $y\rightarrow0^\mp$, while $\rho$ either increases without bound or tends to zero depending again on whether $y$ is positive or negative. This already is a big difference compared with flat space-time in which, due to the property $z(y)=-z(y)$ and to the constant radius $\rho(y)=1$, $y$ and $-y$ were mapped to the same point on the $\rho$-$z$ plane. In this sense, one can think of the extra dimension as opening up or collapsing as an observer attempts to draw closer to the singularity. The fact that both possibilities exist is consistent with the directional nature of this ill-defined region. It can also be observed that $z$ grows extremely faster than the radius and therefore, the mentioned vanishing and infinite limits of the latter cannot be fully appreciated in these numerical solutions. Nevertheless, given a sufficiently small neighborhood of the singularity, one can find solutions with either arbitrarily large or arbitrarily small values of $\rho=I$. 

To aid the reader in correctly visualizing these results, the past profiles can be rearranged in one graphic as follows. For $0<y<1$ we choose to show the negative branch for $dz/dy$ (recall that it appears squared in equation (\ref{dzdysing})), and for $-1<y<0$ we take the positive branch. This is presented in figure \ref{fig:zsing2}. Naturally, moving away from the singularity, the profiles for positive and negative values of $y$ start to approach that of the Minkowski metric. This is due to the first term on the left-hand side of (\ref{dzdysing}) dominating over the second one when $y^2\approx1$. Notice that despite the differential equation being singular at $y=\pm1$, its solution itself, as seen by examining the special case in (\ref{dzdyf}), is regular at those points. Another comment worth highlighting from figure \ref{fig:zsing2} is that the $a$ parameter is not only responsible for curvature (all of the curvature quantities are proportional to a positive power of $a$), but also for the points $y$ and $-y$ being distinguishable in the resulting profiles. Indeed, they represent completely different infinite trajectories to $x\rightarrow0$ depending on the sign of the $y$ coordinate.

Finally, since the profiles $z(\rho)$ obtained here diverge at the singularity, we can associate this geometrical property to the peculiar behavior found in subsection III.C in which geodesics constrained to the throat ($x=0$) take an infinite affine parameter $\lambda$ to meet the singularity. In this five-dimensional analysis we give it the following interpretation. When approaching the singular ring, such observers travel across the fifth dimension, spanning an infinite profile that leads them to the singularity, but that never completely reaches it. If their path is projected into a four-dimensional space-time it would seem as if they are hardly advancing toward the singularity because the effects of the fifth dimension are not being considered. However, by taking into account the extra dimension, it can be realized that the infinite affine parameter needed to reach the singularity is the result of an infinite five-dimensional path. 

Unfortunately, the analysis done here only goes as far as being able to illustrate the five-dimensional geometry near the singularity for the relatively simple $x_0=0$ case, while general geodesics that approach it can travel without being restricted to that plane. These curves, as seen in the past section, also need infinite affine parameter to reach the singular ring. Thus, it is likely that physical observers following said geodesics meet some kind of infinite embedded geometry on their way to the singularity too. One can alternatively consider slices of constant $y_0=0$. However, these are not very interesting due to the fact that $I(x,0)=\left.I_2\right|_{y_0=0}=1$, and therefore, the radius of the fifth dimension is constant. In this case, the solution $z(x)$ for equation (\ref{dzdxi}) is singular at $x=0$, growing without bound. Besides, it was already shown that geodesics attempting to meet the singularity by traveling in the plane $y=0$ were repelled. Consequently, this particular embedded geometry shall not be of any further interest here.

\section{V. Conclusions}

In this paper we have extended a previous study of geodesics in an electromagnetic dipole wormhole, our interest focused mainly on the completeness of this kind of curves. The space-time is a solution to the Einstein-Maxwell equations coupled to either a dilatonic or phantom field and it contains a curvature singularity in the shape of a ring. The singularity is directional in the sense that, depending on the path taken to approach it, curves will encounter either infinite or almost vanishing curvature. As a starting point, special cases of geodesics were examined. The most relevant case to completeness was that of motion constrained to the throat of the wormhole. In the phantom wormhole those geodesics were found to be incomplete due to them being able to meet the singular ring in a finite affine parameter. Numerical analysis was then needed in order to study a set of general geodesics directed toward the singularity. Results revealed that, in contrast to the wormhole with a phantom field and depending on the coupling of the scalar field with the electromagnetic part, it is also possible for geodesics to require an infinite amount of said parameter to reach the ill-defined region of the space-time. Thus, the curves under consideration did not become incomplete. Only wormholes coupled to dilatonic fields exhibit this behavior, physically relevant cases of this are found in a low-energy string theory and in Kaluza-Klein theory. Completeness is mainly achieved due to the directional nature of the singularity.

To better understand the singularity on the dilatonic type of wormholes, we then presented a further analysis of it in the context of the five-dimensional theory by Kaluza and Klein. In such theory, electromagnetism is naturally incorporated into the geometry of space-time through the addition of a fifth dimension. Considering the extra dimension as compactified and periodic, we embedded two-dimensional slices of the space-time in an Euclidean space of dimension three. It was found that, depending on the direction in which an observer approaches the singularity, the radius of the fifth dimension either becomes infinite or tends to zero. Thus, the singularity greatly modifies the geometrical properties of the space-time seen as a five-dimensional object. The embedded profiles also became infinite due to the singularity, giving a possible explanation to the previously described feature of geodesics needing infinite affine parameter to reach it. The infinite profiles were interpreted as endless paths that lead an observer to the singular region without ever meeting it.

Finally, these dilatonic wormholes constitute examples in which the presence of a curvature singularity does not necessarily imply geodesic incompleteness, leaving opened the possibility of them being space-times with complete causal curves. To prove this final characteristic, though, the effects of the singularity on time-like curves of bounded acceleration must be studied too. If, on the contrary, other non-geodesic curves are shown to reach the singularity, then we are left with a particular and interesting instance of a singular space-time despite it being geodesically complete. \\

 \textbf{Acknowledgments.} This work was partially supported by CONACyT M\'exico under grants CB-2011 No. 166212, CB-2014-01 No. 240512, Project No. 269652, Fronteras Project 281, and grant No. I0101/131/07 C-234/07 of the Instituto Avanzado de Cosmolog\'ia (IAC) collaboration (http://www.iac.edu.mx/). J.C.D.A. acknowledges financial support from CONACyT postdoctoral fellowships too. The authors would also like to thank Hugo A. Morales-T\'ecotl for his support during the writing process of this paper. \\

\end{document}